\documentclass[twocolumn]{aastex701}

\newcommand{\e}[1]{\times 10^{#1}}
\newcommand{\msun}{\rm{M_\odot}}

\usepackage{graphicx}	% Including figure files
\usepackage{xcolor}
\usepackage{amsmath}	% Advanced maths commands
\usepackage{multirow}
\usepackage{enumitem}
\usepackage{siunitx}

\definecolor{xlinkcolor}{cmyk}{1,1,0,0}

\shorttitle{GGSL Host and Line of Sight Dependence}
\shortauthors{Roche et al.}

\begin{document}

\title{Host Dependence and Line-of-Sight Effects on Galaxy-Galaxy Strong Lensing in Clusters}

% ------------ AUTHORS ------------

\author[0000-0002-3400-6991]{Cian Roche}
\affil{Department of Physics and Kavli Institute for Astrophysics and Space Research,\\Massachusetts Institute of Technology, Cambridge, MA 02139, USA}
\email[show]{roche@mit.edu}

\author[0000-0001-5226-8349]{Michael McDonald}
\affil{Department of Physics and Kavli Institute for Astrophysics and Space Research,\\Massachusetts Institute of Technology, Cambridge, MA 02139, USA}
\email{mcdonald@space.mit.edu}

\author[0000-0002-5554-8896]{Priyamvada Natarajan}
\affiliation{Department of Physics, Yale University, New Haven, CT 06511, USA}
\affiliation{Department of Astronomy, Yale University, New Haven, CT 06511, USA}
\email{priyamvada.natarajan@yale.edu}

\author[0000-0002-7559-0864]{Keren Sharon}
\affiliation{Department of Astronomy, University of Michigan, 1085 South University Avenue, Ann Arbor, MI 48109, USA}
\email{kerens@umich.edu}

\author[0000-0001-7040-4930]{Isaque Dutra}
\affiliation{Department of Physics, Yale University, New Haven, CT 06511, USA}
\email{isaque.dutra@yale.edu}

\author[0000-0003-1225-7084]{Massimo Meneghetti}
\affiliation{INAF-Osservatorio di Astrofisica e Scienza dello Spazio di Bologna, 
Via Piero Gobetti 93/3, 40129 Bologna, Italy}
\email{massimo.meneghetti@inaf.it}

\author[0000-0001-8593-7692]{Mark Vogelsberger}
\affil{Department of Physics and Kavli Institute for Astrophysics and Space Research,\\Massachusetts Institute of Technology, Cambridge, MA 02139, USA}
\affil{Fachbereich Physik, Philipps Universit\"at Marburg, D-35032 Marburg, Germany}
\email{mvogelsb@mit.edu}

% ------------- ABSTRACT ------------
\begin{abstract}
The cross section for galaxy-galaxy strong lensing (GGSL) events in galaxy clusters has repeatedly been found to be higher in observations than in cosmological simulations. We revisit this discrepancy using updated simulation methodology and investigate the dependence of the GGSL probability, $P_{\rm GGSL}$, on host-cluster lensing properties, baryonic physics, and correlated and uncorrelated line-of-sight structure. We find that correlated material within $\sim 35$ cMpc of the cluster along the line of sight enhances $P_{\rm GGSL}$ by a few percent for typical systems and by up to $\sim15\%$ for the most efficient lenses. At fixed cluster mass, dark-matter-only simulations yield GGSL probabilities up to an order of magnitude lower than hydrodynamical simulations. We also find a strong dependence on the host-cluster Einstein radius, with an approximate scaling $P_{\rm GGSL} \propto \theta_{\rm E}^{2}$. Matching simulated and observed clusters in both mass and Einstein radius seems to reduce the discrepancy relative to previous comparisons. However, our analysis does not clearly resolve the GGSL discrepancy, as the inferred tension depends strongly on the field-of-view definition: square fields are approximately consistent with simulations, while cluster member-bounded fields yield observed probabilities a factor of $\sim 2$--$3$ higher. Until observations and simulations share a matched selection function and matched measurement methodology, the residual tension cannot be cleanly attributed to either astrophysics or cosmology. 
\end{abstract}


% ------------ BODY ------------
\section{Introduction}\label{sec:intro}
Galaxy clusters have proven to be a valuable testing ground for dark matter physics. They provided the first compelling observational evidence for an unseen and dominant matter component \citep{zwicky1933,zwicky:1933:ComaDM}, have acted as a powerful probe of dark matter self-interaction properties in merging systems \citep{randall:2008:bullet,robertson2017:bullet_sidm}, and as gravitational lenses they have been used to constrain the mass distribution of clusters to high precision without assuming dynamic equilibrium \citep{Natarajan:2024, cerny2025:mass_profiles}. In addition, they offer a unique opportunity to study the objects behind these lenses \citep{zitrin2026:lensing_jwst_review}. The field of cluster strong lensing has seen significant advances in observational resolution and depth in recent years, with Hubble Space Telescope (HST) programs such as CLASH \citep{postman2012:CLASH_initial}, HST Frontier Fields \citep{lotz2017:HFF}, GLASS \citep{schmidt2014:GLASS, treu2015:glass}, and RELICS \citep{coe2019:RELICS}. This progress has continued with James Webb Space Telescope (JWST) programs such as GLIMPSE \citep{atek2025:GLIMPSE}, SLICE \citep{cerny2026:slice}, VENUS \citep{fujimoto2025:VENUS} and UNCOVER \citep{bezanson2024:uncover}, pushing the number of observed multiple image constraints for a single cluster to as many as $\sim300$ \citep{rihtarsic2025:macs_j0416_strong_lens_model}. This leap in depth and constraining power has enabled the study of fainter and more distant objects than ever before \citep{welch2022:earendel, boylan_kolchin2023:high_z_stellar_mass_lcdm_consraint, haojing2023:macs_j0416_transients, atek2023:high_z_magnified_candidates} and has allowed for a deeper interrogation of the properties of dark matter halos on small scales \citep{keeley2024:jwst_lensed_quasars, nierenberg2024:jwst_lensed_quasars,  ephremidze2025:dm_substructure_detection_systematics, gannon2025:dm_substructure_stripping,  perera2026:dm_substructure_astrometric_asymmetry, gilman2026:dm_free_streaming, natarajan+2026ApJL}.\\

Recently, a tension has emerged between observations and simulations with regard to the lensing properties of member galaxies in clusters. In particular, the total cross section for galaxy-galaxy strong lensing (GGSL) events within clusters has been identified as significantly higher in observations than in simulated clusters, potentially implying that cluster members in simulations are significantly less compact than in reality or that the distribution in mass or space of subhalos in simulated clusters is not consistent with observed clusters \citep{natarajan+2017}. This was first identified in \cite{meneghetti2020} (hereafter M20), after which the dependence of the GGSL probability on simulation resolution and feedback schemes was identified as a potential solution to the reported tension \citep{robertson2021:ggsl, bahe2021:ggsl}. However, follow-up studies found that improving simulation resolution could not resolve the discrepancy, and that while the feedback scheme could make a significant difference to the simulated cross section, adopting feedback that significantly reduces the GGSL tension introduces other tensions with observations, such as the introduction of very large, bright galaxies not seen in observations \citep{ragagnin2022:ggsl, meneghetti2022, meneghetti2023}. It was also observed that the redistribution of mass within the Einstein radius in a manner consistent with cold dark matter (CDM) is insufficient to explain the difference between observations and simulations (\citealt{tokayer2024}; T24). These facts have been interpreted as a direct challenge to the standard cosmological paradigm of CDM with a cosmological constant $\Lambda$ \citep{natarajan+2026ApJL}. It has been demonstrated in \cite{dutra2025:ggsl} that sufficiently steep inner density slopes could resolve the tension, but such slopes are inconsistent with CDM, even with baryonic contributions, and thus could signal a scenario consistent with steeper slopes characteristic of core-collapse self-interacting dark matter (SIDM) in the observations. The strong lensing efficiency of individual cluster member galaxies can be understood as arising from a combination of intrinsic properties and additional boost provided by the large-scale cluster environment, and disentangling these effects is important for understanding the underlying dark matter physics. This will be studied in detail in \href{Dutra et al. 2026}{Dutra et al. 2026} (in prep).\\

Some features of this tension remain to be understood, such as the exact effect of mismatched selection functions and the methodology used to determine the mass distribution in observations and simulations. The observational GGSL probability estimates are subject to a selection function, and the mass distribution of each observed cluster is reconstructed by modeling the gravitational lensing potential from observed strong- and weak-lensing evidence, which is subject to gravitational lens modeling choices and the availability of constraints. In contrast, in simulations, one has access to 3D position information for all star particles, dark matter particles, and gas cells, and the projected mass distribution can be computed directly. Furthermore, limited simulation resolution and gravitational softening limit the interpretability of the mass distribution below certain mass and length scales \citep{power:2003:convergence, bosch:2018:convergence, roche:2024}. While observations unavoidably include the effect of all material along the line of sight, not only the cluster of interest, in simulations it is common to extract only the gravitationally bound massive object identified as the cluster to be used as the primary lens. The effect of line of sight material has been estimated preliminarily in M20, where it was found to contribute insignificantly to the cross section, but this test was limited to sampled analytic profiles in an assumed unbiased sample of the universe, not accounting for correlations between structures. \\

\begin{deluxetable*}{ccccccccc}
\tabletypesize{\footnotesize} 
% \tablewidth{\textwidth}
\tablecaption{\normalfont\raggedright Observed clusters used to determine the GGSL probability. Parameters are as follows, from left to right: (1) the object identifier, (2) the cluster redshift $z_\ell$, (3) the redshift range used to define cluster membership $\Delta z_\ell$, (4) the effective Einstein radius $\theta_E$, (5) the mass definition used to report the inferred cluster mass, (6) the cluster mass, where ${}^{\dagger}$ means the estimate comes from a different strong lens model than the one used to measure the GGSL probability, (7) the GGSL probability with a \qty{200}{\arcsecond} square field of view \citep{tokayer2024}, (8) the GGSL probability with the bounding curve field of view, with median square root area of $\sim\qty{160}{\arcsecond}$ \citep{meneghetti2023}. (9) Reference for the lens model. A source plane redshift of $z_s=6$ is used for columns 4, 7, and 8. GGSL probabilities are reported in units of $10^{-6}$.\label{tab:obs}}
\tablehead{
\colhead{Object} & 
\colhead{$z_\ell$} &  
\colhead{$\Delta z_\ell$} &
\colhead{$\theta_E\,[\qty{}{\arcsecond}]$} & 
\colhead{Mass Type} & 
\colhead{Mass $[10^{15}\,\msun]$} & 
\colhead{$P_{\rm GGSL, \square}$} &
\colhead{$P_{\rm GGSL, b}$} &
\colhead{Model}%\vspace{0.3em}
}
\startdata
Abell 2744 & 0.308 & 0.040 & $26.03^{+0.29}_{-0.11}$ & $M_{\rm enc}(1.3\,{\rm Mpc})$ & 2.2 & 856 & $1782^{+537}_{-347}$ & \cite{mahler2018:sl_model}\\
Abell S1063  & 0.346 & 0.027 & $36.79^{+0.34}_{-0.45}$ & $M_{200c}$ & $2.03 \pm 0.67$ & 361 & $1772^{+713}_{-396}$ & \cite{bergamini2019:HFF}\\
MACS J0416 & 0.397 & 0.028 & $31.31^{+0.22}_{-0.35}$ & $M_{200c}{}^{\dagger}$ & $1.04\pm0.22$ & 343 & $1449^{+648}_{-263}$ & \cite{bergamini2023:macs_j0416}\\
MACS J1206 & 0.439 & 0.028 & $33.19^{+0.28}_{-0.15}$ & $M_{200c}$ & $1.59 \pm 0.36$ & 1168 & $5456^{+992}_{-1073}$ & \cite{bergamini2019:HFF}\\
PSZ1 G311 & 0.443 & 0.024 & $34.69^{+0.69}_{-0.02}$ & $M_{\rm enc}(200\,{\rm kpc})$ & $0.200\pm 0.001$ & 452 & $3074^{+2313}_{-210}$ & \cite{pignataro2021:PSZ1G311_lens_model}
\enddata
\end{deluxetable*}

To resolve these problems, it is in principle necessary to produce strong lensing mocks from simulated clusters with properties as comparable as possible to the observed clusters, including the effect of all line of sight structure, and to lens model those mocks in precisely the same manner as the observations. This requires significant lens modeling effort, or an automated lens modeling pipeline for the simulations that does not yet exist, and would in any case not match the procedure used for modeling observations previously used to study the GGSL tension. In this paper, we attempt an updated estimate of the line-of-sight effects using light cones generated from cosmological simulations. We extract information directly from these light cones rather than inferring the lensing properties via lens modeling. We also study the dependence of the GGSL probability on the mass and Einstein radius of the host cluster, and test the status of the tension with a sample of simulated clusters matched in both of these parameters to observed clusters which have been used to study the GGSL tension in the literature.\\ 

The remainder of this paper is structured as follows: in Sec.~\ref{sec:data} we outline the sample of observed clusters (Sec.~\ref{sec:obs_data}) and simulated data (Sec.~\ref{sec:sim_data}) we will use to study the GGSL discrepancy. In Sec.~\ref{sec:methods} we outline the definition and computation of the GGSL probability, and in Sec.~\ref{sec:results} we examine the dependence of that probability on the properties of the cluster (Sec.~\ref{sec:results_primary}), correlated and uncorrelated structure along the line of sight (Sec.~\ref{sec:results_los}), and the inclusion or exclusion of baryons (Sec.~\ref{sec:results_baryons}). We then determine the status of the tension in this data in Sec.~\ref{sec:ggsl_comparison}. We conclude in Sec.~\ref{sec:conclusions} with recommendations for future work that will aid in mitigating systematics and establishing a more robust measure of the GGSL discrepancy. We use the cosmological parameters of \cite{planck2016} for all computations, unless otherwise stated.

\section{Data}\label{sec:data}
\subsection{Observations}\label{sec:obs_data}
For our observational comparisons, we include the GGSL probability measurements of \cite{meneghetti2023} and \cite{tokayer2024}\footnote{We use the dual-Pseudo-Isothermal-Elliptical models from this paper, as the truncated NFW results are intended to boost the GGSL probability.}. A part of this sample comprises the ``reference sample'' of M20 -- MACS J1206.2-0847 (MACS J1206), MACS J0416.1-2403 (MACS J0416), and Abell S1063 -- which were selected on the basis of having well-constrained lens models and available stellar kinematics of galaxy cluster members. In addition to the M20 reference sample objects, Abell 2744 and PSZ1-G311.65-18.48 (PSZ1 G311) are included, again with stellar kinematics data to constrain the properties of the cluster members in the lens models. In addition to stellar kinematics these clusters were also selected for possessing large amounts of lensing evidence, which is related to several properties of the clusters including their mass, elongation along the line of sight, and their concentration \citep{Natarajan:2024}. This results in a complicated selection function that we do not attempt to reproduce in the simulations; future work utilizing a sample of observed cluster lens models with a well-understood selection function, such as the sample of SPT clusters observed by the \cite{sptsnap} HST/SNAP program \citep{remolina2021:HST_SNAP} or the SLICE survey \citep{SLICE, cerny2026:slice}, would significantly improve the reliability of the observation-simulation comparison. \\

We collect the properties of the observed sample in Table~\ref{tab:obs}, which contains much of the information of Table~1 in \cite{meneghetti2023} and Table~1 in \cite{tokayer2024}. The mass of MACS J0416 has been quoted as $M_{132c}=(1.3\pm0.3)\times 10^{15} \,{\rm M_\odot}$ by \cite{umetsu2016:CLASH}, as noted in \cite{tokayer2024}, but here we report the \cite{bergamini2019:HFF} model mass because an $M_{200c}$ estimate is available for that model. Additionally, \cite{mahler2018:sl_model} does not report uncertainties on the enclosed mass for Abell 2744, but \cite{jauzac2016:Abell2744_strong_weak} determine with weak lensing that the mass enclosed within the same radius is $M_{\rm enc}(1.3\,{\rm Mpc})=(2.3\pm 0.1)\times 10^{15}\,{\rm M_\odot}$. All observed mass models used here are parametric and have been constrained using \texttt{Lenstool} \citep{jullo2007:lenstool, kneib2011:lenstool_code}, an open-source code commonly used to invert the lens equation for cluster lenses.

\subsection{Simulation Data}\label{sec:sim_data}

For the theoretical predictions, we make use of two suites of cosmological simulations with different and complementary strengths: one with a large number of massive cluster zoom-in simulations, and another with a uniform box that can be used to construct light cones. We describe each set of simulation data below.

\subsubsection{TNG-Cluster}
 For generating a large number of massive cluster lens planes \textit{without} including correlated or uncorrelated structure along the line of sight, we use the suite of cluster zoom-in simulations TNG-Cluster \citep{nelson2024:TNG-Cluster}. These simulations were initially run in a low-resolution, dark matter only box with side length ($1\,{\rm Gpc}$), in which 352 massive cluster targets were identified. These regions were then re-simulated with higher resolution and the IllustrisTNG galaxy formation model \citep{pillepich:2017:TNGgalaxyformation}. The resulting clusters span a mass range from roughly $10^{14}\,{\rm M_\odot}$ to $10^{15.5}\,{\rm M_\odot}$, but with a mass function weighted toward larger objects. These objects are conducive to the present analysis, as the observed clusters we will compare to are very massive or elongated along the line of sight, resulting in a large Einstein radius (Table~\ref{tab:obs}). The mass resolution of the dark matter particles in TNG-Cluster is $6.1\times 10^7\,{\rm M_\odot}$ and the mean baryonic cell mass is $1.2\times 10^7\,{\rm M_\odot}$, with a gravitational softening of $1.48\,{\rm kpc}$ for the dark matter and star particles, with a minimum adaptive softening for the gas of $370\,{\rm pc}$. For the results discussed in Sec.~\ref{sec:results}, we select only those groups for which $M_{200c}>5\times 10^{14}\,{\rm M_\odot}$ at the redshift of the chosen snapshot (which always matches the redshift of the primary lens in lensing computations $z_\ell$). This results in 142 objects at $z_\ell=0.38$.

\subsubsection{IllustrisTNG}
The TNG-Cluster simulations are not well-suited for testing the impact of correlated and uncorrelated structure, since the material outside of re-simulation regions is low-resolution and runs without hydrodynamics. Therefore, for these tests, we instead make use of the uniform hydrodynamical simulations on which TNG-Cluster is based, namely IllustrisTNG (TNG; \citealt{springel:2017:ITNG, pillepich:2017:ITNG, marinacci:2018:ITNG, naiman:2018:ITNG, nelson2021:illustristng}). We construct light cones from the TNG300-1 box, which is the highest-resolution variant of the largest TNG box and therefore contains the most relevant cluster lens objects for comparison with our observations, but notably contains significantly fewer massive clusters than TNG-Cluster. TNG300-1 was run in a $205\,{\rm Mpc}\,h^{-1}$ box, uses $2500^3$ dark matter particles and an identical initial number of gas particles, a baryonic mass resolution of  $7.6\e{6}\,\msun\,$$h^{-1}$ and DM mass resolution of $4.0\e{7}\,\msun\,$$h^{-1}$. The gravitational softening length of the stars and DM is $1\,{\rm kpc}\,h^{-1}$ at $z=0$ and the minimum adaptive gas softening length is $0.25\,{\rm ckpc}\,h^{-1}$.\\

We construct the TNG300-1 light cones according to the simulated cluster strong lensing pipeline described in \cite{roche2026} (hereafter R26). This method uses all snapshots of a parent simulation between some minimum and maximum redshifts $z_{\rm min}$ and $z_{\rm max}$ to generate source and lens planes along a line of sight, with a large group (cluster) at some redshift $z_\ell$, chosen to match observed strong lensing cluster redshifts. When discussing the dependence of GGSL on the inclusion of correlated structure, the midpoint in redshift between the snapshot at $z_\ell$ and its neighboring snapshots are used as bounds along the line of sight, resulting in $\Delta_z \simeq 0.02$ and a corresponding thickness in comoving distance of $\sim \pm 35\,{\rm cMpc}$ in these tests. When computing GGSL in TNG, we select only those groups with $M_{200c}>3\times 10^{14}\,{\rm M_\odot}$, a lower cut than used for TNG-Cluster. This results in 18 objects at $z_\ell=0.38$.\\

The cosmology used in both TNG and TNG-Cluster is that of \citealt{planck2016} with parameters $\Omega_{\rm{m}} = \Omega_{\rm{CDM}} + \Omega_{\rm{b}} = 0.3089,\: \Omega_{\rm{b}} = 0.0486,\: \Omega_\Lambda = 0.6911,\: \sigma_8 = 0.8159,\:n_s = 0.9667,\: h = 0.6774$. All mass planes are computed from the simulation particles using the plane creation pipeline described in R26, which involves smoothing particles of a given species according to the local density of that species onto a grid. We use the fiducial axis ratios for the box remapping, as described in R26, resulting in $z_{\rm max}=4.2$ for the light cones. We include only groups with a mass of at least 200 times the dark matter particle mass of the simulation in the mass planes, and we use 64 nearest neighbors for the smoothing length computation for dark matter, 32 neighbors for star particles, and the smoothing lengths for gas cells are computed from the cell volumes $V_{\rm cell}$ as $(3V_{\rm cell}/4\pi)^{1/3}$.

\section{Methodology}\label{sec:methods}
\subsection{Lensing Formalism}
We first review the formalism we will use to describe the lensing variables. See \cite{meneghetti2022:lensing_book} for a more detailed overview. Given an assumed ``thin'' lens at redshift $z_\ell$ with surface mass density $\Sigma$, and a source plane at redshift $z_s$, one may compute the dimensionless surface mass density, or ``convergence''
\begin{equation}
    \kappa = \frac{\Sigma(\vec \theta)}{\Sigma_{crit}}, \quad \text{where} \quad \Sigma_{crit} = \frac{c^2}{4\pi G} \frac{D_s}{D_\ell D_{\ell s}}
\end{equation}
which depends on the vector position on the image plane $\vec \theta = (\theta_1, \theta_2)$ and is defined in terms of the cosmology-dependent angular diameter distances between the observer and source ($D_s$), the observer and lens ($D_\ell$), and the lens and source ($D_{\ell s}$). The lensing potential $\psi$ then satisfies the Poisson equation 
\begin{equation}
    \nabla^2 \psi = 2\kappa(\vec \theta).
\end{equation}
The Jacobian $\mathcal{A}$ for the transformation from points on the source plane $\vec \beta = (\beta_1, \beta_2)$ to points on the image plane can be written as
\begin{equation}
    \mathcal{A}_{ij} = \frac{\partial \beta_i}{\partial \theta_j} = \delta_{ij} - \frac{\partial^2\psi}{\partial \theta_i \partial\theta_j}.
\end{equation}
The convergence is recovered via the trace of the Jacobian, since ${\rm tr}(\mathcal{A})=2-\nabla^2\psi = 2(1-\kappa)$, and the ``shear'' tensor is the trace-free part $S = \mathcal{A} - \frac{1}{2}{\rm tr}(\mathcal{A})\mathbb{I}_2$ where $\mathbb{I}_2$ is the identity matrix in 2 dimensions. This can be expressed in matrix form as 
$$
S = 
\begin{pmatrix}
-\gamma_1(\vec\theta) & -\gamma_2(\vec\theta) \\
-\gamma_2(\vec\theta) & \gamma_1(\vec\theta)
\end{pmatrix}
$$
where
\begin{align}
\gamma_1(\vec\theta) &:= \frac{1}{2}\left(\frac{\partial^2\psi}{\partial \theta_1^2} - \frac{\partial^2\psi}{\partial \theta_2^2}\right) \\
\gamma_2(\vec\theta) &:= \frac{\partial^2\psi}{\partial \theta_1 \partial \theta_2} = \frac{\partial^2\psi}{\partial \theta_2 \partial \theta_1}
\end{align}
If we then define the shear magnitude $\gamma = \sqrt{\gamma_1^2 + \gamma_2^2}$ and rewrite the Jacobian as
\begin{equation}
  \mathcal{A}=  
  \begin{pmatrix}
1-\kappa-\gamma_1 & -\gamma_2 \\
-\gamma_2 & 1-\kappa + \gamma_1
\end{pmatrix},
\end{equation}
we can compute the magnification as 
\begin{equation}
    \mu = \frac{1}{{\rm det}\mathcal{A}} = \frac{1}{(1-\kappa)^2-\gamma^2}
\end{equation}
We can now identify the ``critical curves'' in the image plane where the magnification diverges, i.e. the inverse magnification is zero 
\begin{equation}
    \mu^{-1} = (1-\kappa-\gamma)(1-\kappa + \gamma) = 0
\end{equation}
which occurs where $\kappa+\gamma = 1$ (the ``tangential'' critical curves) or $\kappa-\gamma = 1$ (the ``radial'' critical curves). For a smooth isolated lens, the condition to be ``supercritical'' (that is, to produce critical curves, and hence caustics and multiple images) is that there exists a region of nonzero area within which $\mu < 0$ \citep[Sec.~5.4.3]{schneider1992:lensing_book}.\footnote{Intuitively, the inverse magnification $\mu^{-1} = \det A$ is a smooth function that tends to unity far from the lens and vanishes on the critical curves. If $\mu^{-1} < 0$ on some region, continuity therefore guarantees that this region is bounded by a locus $\mu^{-1} = 0$, i.e.\ by a critical curve.} Mapping each of these curves back to the source plane produces the so-called ``caustics'' for this primary lens. We define the ``effective Einstein radius'' of a tangential critical curve with area $A_{\rm crit}$ in the image plane as 
\begin{equation}
    \theta_E = \sqrt{\frac{A_{\rm crit}}{\pi}},
\end{equation}
which is interpretable as an exact radius in the case of axially symmetric mass distributions. \\

Generalizing this to the case where matter is distributed at multiple redshifts typically involves the approximation of \textit{multi-plane lensing}, in which the deflection of light at any given plane depends on the deflections at all planes before it. We will not describe the formalism here, but instead direct the reader to other previously published resources \citep{schneider1992:lensing_book,petkova2014:glamer2}. In this paper, we will frequently discuss both single-plane lenses and multi-plane lenses with $\sim 70$ planes distributed throughout a light cone. 

\subsection{The Galaxy-Galaxy Strong Lensing Probability}
The lensing efficiency of member galaxies in galaxy clusters depends on several factors, such as the shape of their density profile, the properties of their environment, and the redshifts of the cluster and source planes. The shape of the inner density profile in the cores of galaxies is a powerful probe of the properties of dark matter \citep{NFW:1996, deblok2010:core_cusp, bull:2016, bullock2017:small_scale_lcdm_challenges}, and strong gravitational lensing has proven a valuable tool in studying the mass distribution of galaxies on $\lesssim 10\,{\rm kpc}$ scales \citep{treu2010:sl_review, shajib2024:sl_galaxies_review}. The galaxy-galaxy strong lensing probability is an aggregate metric of the lensing power of all galaxy-scale members of a galaxy cluster, defined as follows. \\

Once the critical curves of a cluster lens have been identified (for some source plane redshift $z_s$), to be consistent with the literature we consider only those critical curves with an effective Einstein radius \qty{0.5}{\arcsec}$< \theta_E <$\qty{3.0}{\arcsec}, chosen to select for galaxy cluster members. The upper limit excludes large merging/infalling groups, or very large member galaxies produced in the simulations which do not appear to have a counterpart in the observations and dominate the GGSL signal (\citealt{meneghetti2022, meneghetti2023}; M22, M23). We examine the presence of large-$\theta_E$ critical curves in the observations and simulations considered here in Sec.~\ref{sec:results}. The lower limit is used to ensure that included objects and their corresponding curves are well-resolved in simulations, and well-constrained in observations. We denote the areas of the corresponding caustic curves as $\sigma_i$. We then define the galaxy-galaxy strong lensing cross section for this cluster to be 
\begin{equation}\label{eq:GGSL}
    \sigma_{\rm GGSL} := \sum_{i} \sigma_i, \quad \qty{0.5}{\arcsecond} < \theta_{E,i} < \qty{3.0}{\arcsecond}.
\end{equation}

The corresponding GGSL probability is obtained by normalizing this summed source plane area by the area of the field of view, mapped back to the source plane. 
\begin{equation}\label{eq:pggsl}
    P_{\rm GGSL} = \frac{\sigma_{\rm GGSL}}{\sigma_{\rm FOV}}.
\end{equation}
In recent works, the source plane field of view $\sigma_{\rm FOV}$ is computed differently in observations and simulations \citep{meneghetti2022, meneghetti2023}. In simulations, the field of view used to compute the mass distribution is a square; however, in observations, the mass distribution is only constrained to the modeled region, approximately described by a curve enclosing the outermost cluster member galaxies included in the lens model. To reflect the fact that any galaxy-scale lenses outside this region are not explicitly modeled and thus would not contribute to the GGSL probability, the area of the source-plane counterpart to the curve bounding the outermost member galaxies is used for $\sigma_{\rm FOV}$. \\

\begin{figure*}
    \centering
    \includegraphics[width=1.0\linewidth]{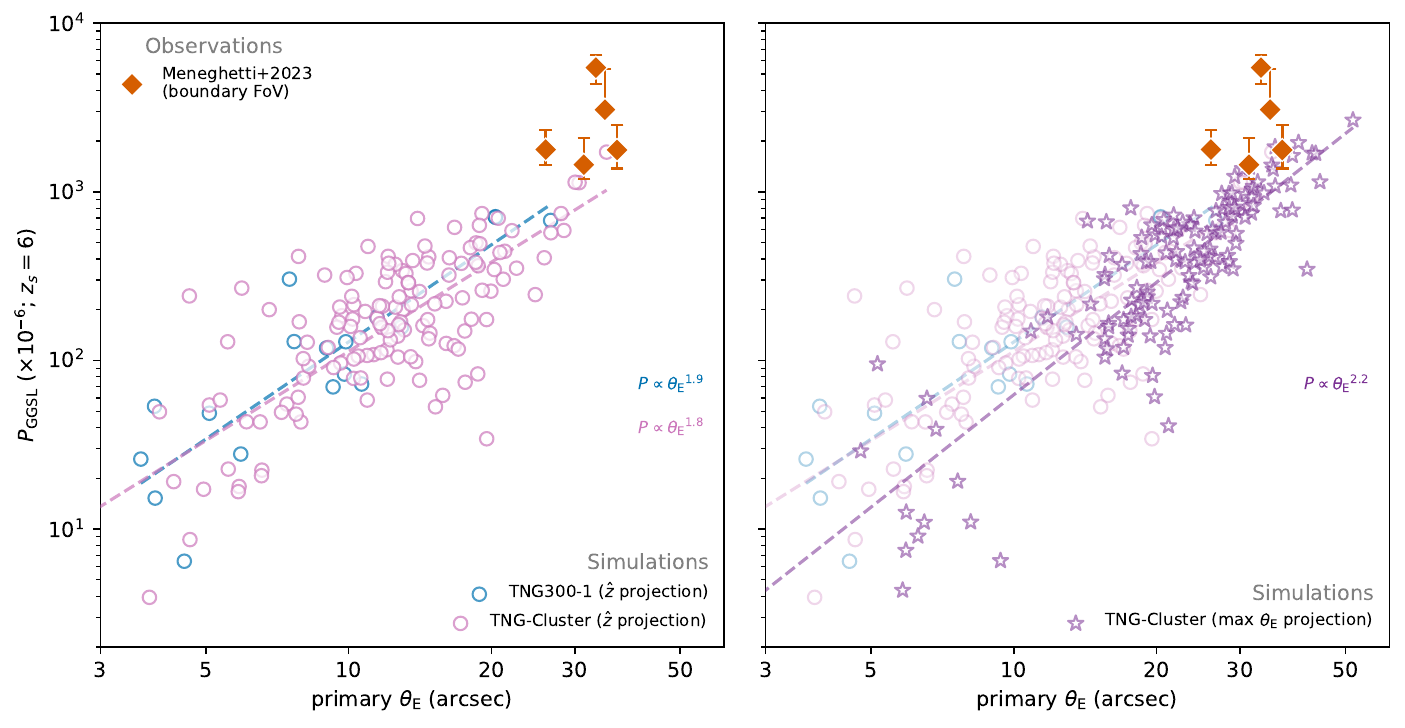}
    \caption{ The Galaxy-Galaxy Strong Lensing (GGSL) probability in observations and simulations as a function of the primary lens effective Einstein radius. Simulations use only the friends-of-friends identified group as the lens with no correlated structure, and all simulated lenses are placed at $z_\ell=0.38$. Power law fits are assumed to have the form $P_{\rm GGSL}(\theta_E) \propto (\theta_E/\,\qty{10}{\arcsecond})^b$ and are fit with least squares in log space. Simulations use a \qty{150}{\arcsecond} square field of view. Orange diamonds are observations which utilize a curve bounding the outermost cluster members in the strong lens model as the field of view. (\emph{Left}) TNG300-1 groups with $M_{200c}>3\times 10^{14}\,{\rm M_\odot}$ and TNG-Cluster groups with $M_{200c}>5\times 10^{14}\,{\rm M_\odot}$, each projected along $\hat{z}$. (\emph{Right}) The same, additionally showing the TNG-Cluster data projected along axes optimizing $\theta_E$.}
    \label{fig:mass_dep}
\end{figure*}

This mismatch in definitions has the potential to introduce a bias toward higher GGSL probabilities in the observations, as only the reconstructed cluster area is used in the computation, whereas in simulations the whole field of view is used regardless of the projected shape of the cluster. A highly elongated cluster, for example, could ``miss'' critical curves near the ends while including much ``empty'' space in a manner that would not appear in the observations. Any extra space included in the simulation fields of view that does not contain any subhalos will artificially reduce $P_{\rm GGSL}$ relative to the observations. As a result, in Sec.~\ref{sec:results} we will examine GGSL probabilities for the observed clusters determined using a bounding-curve field of view definition (as in \cite{meneghetti2023}), and using a square field of view (as in \cite{tokayer2024}), noting that some value between both estimates, but closer to the boundary curve result, is likely the most appropriate comparison to simulation data. Of course, performing strong lens modeling on simulated images and reconstructing the GGSL probability in precisely the same way as the observations is preferable, but we leave this for future work.

\begin{figure}
    \centering
    \includegraphics[width=0.9\linewidth]{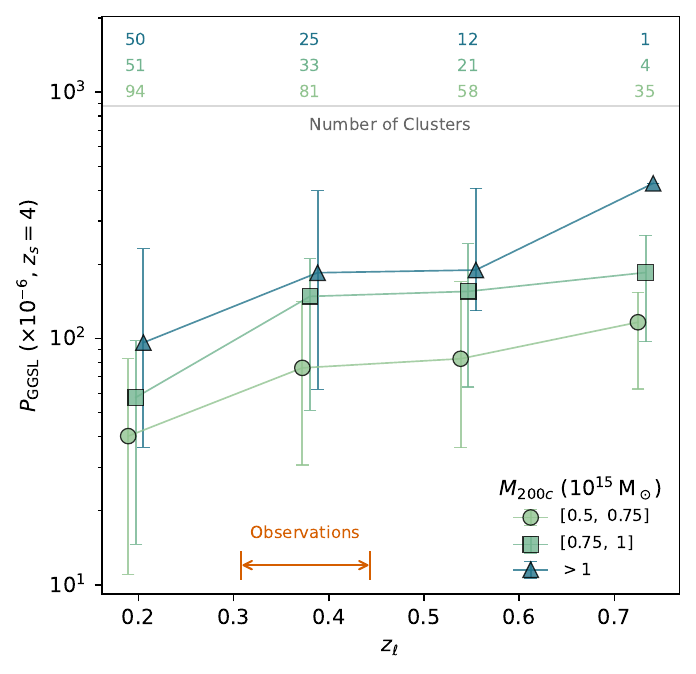}
    \caption{Evolution of the Galaxy-Galaxy Strong Lensing (GGSL) probability with the redshift of the primary lens cluster, binned by cluster mass. Points are medians and errorbars shown are 95\% confidence intervals, and quantities in the top panel reflect the number of clusters at that redshift in that mass bin used to compute the errorbar. The range of lens redshifts for the observations used here is annotated in orange. A square field of view of side length \qty{200}{\arcsecond} was used. The median probability changes by $\sim 15\%$ over the window in $z_\ell$ defined by the observations.}
    \label{fig:sigma_vs_zl}
\end{figure}

\begin{figure*}
    \centering
    \includegraphics[width=1.0\linewidth]{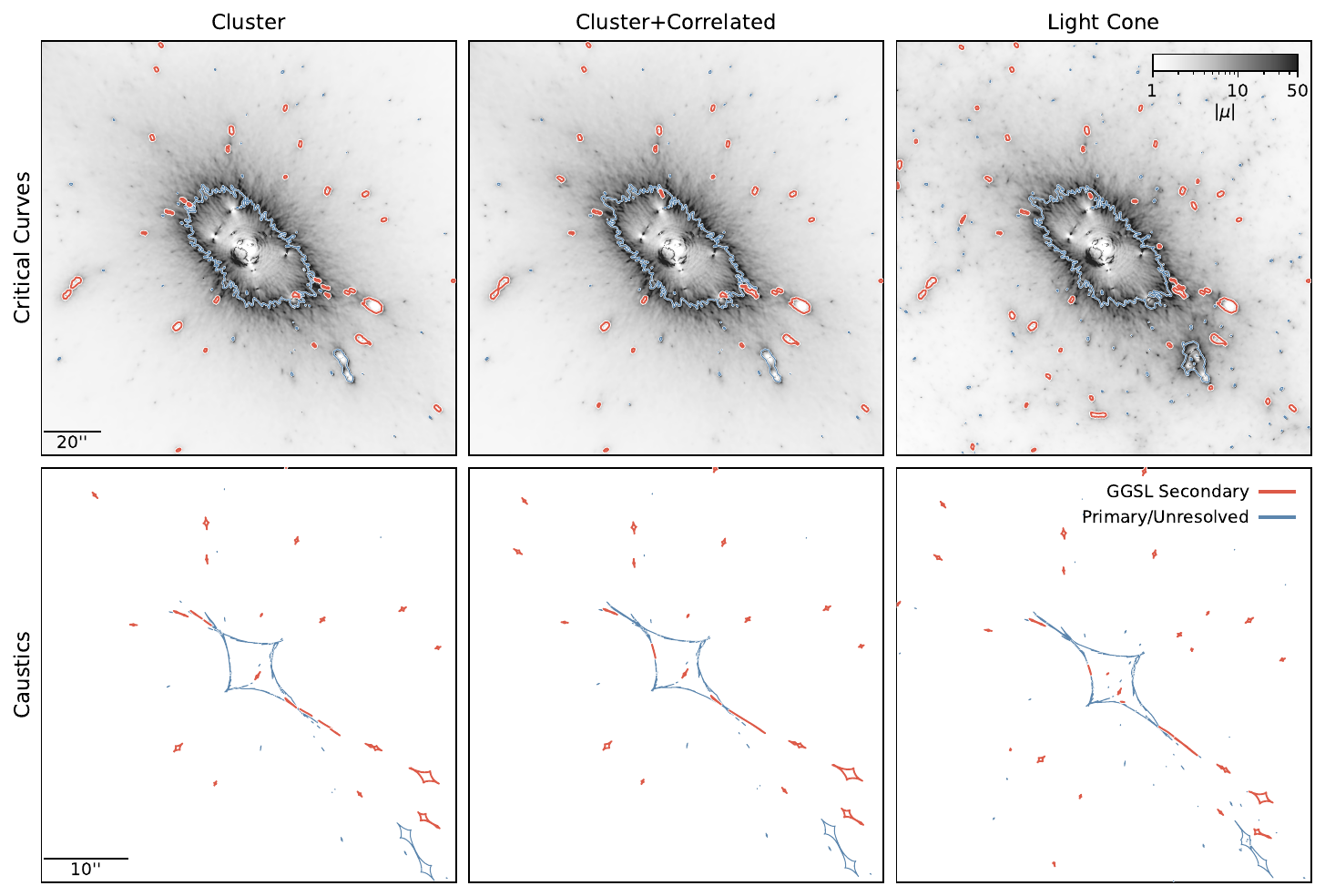}
    \caption{Comparison of the critical curve and caustic structure for a simulated cluster only, that simulated cluster and the surrounding correlated structure, and a full light cone with multi-plane ray tracing. (\emph{Top row}) The critical curves are overlaid on the magnification map, each with source plane redshift $z_s=4.0$ in a \qty{150}{\arcsecond} field of view. (\emph{Bottom row}) The caustics are shown in the source plane in a \qty{50}{\arcsecond} field of view. The primary lens is at $z_\ell=0.38$, and the mass map is computed using the particle data for Group 0 in the TNG300-1 redshift $0.38$ group catalog.}
    \label{fig:ggsl_los_dep}
\end{figure*}

\begin{figure*}
    \centering
    \includegraphics[width=0.45\linewidth]{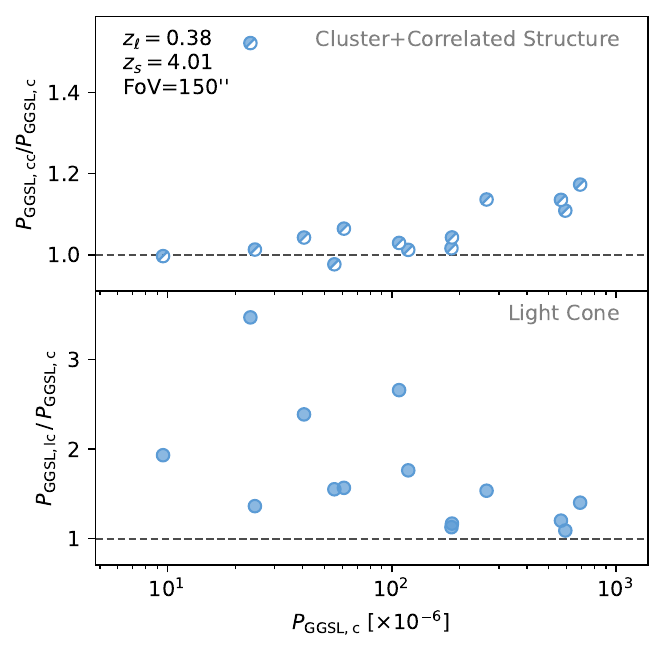}\hspace{2em}
    \includegraphics[width=0.45\linewidth]{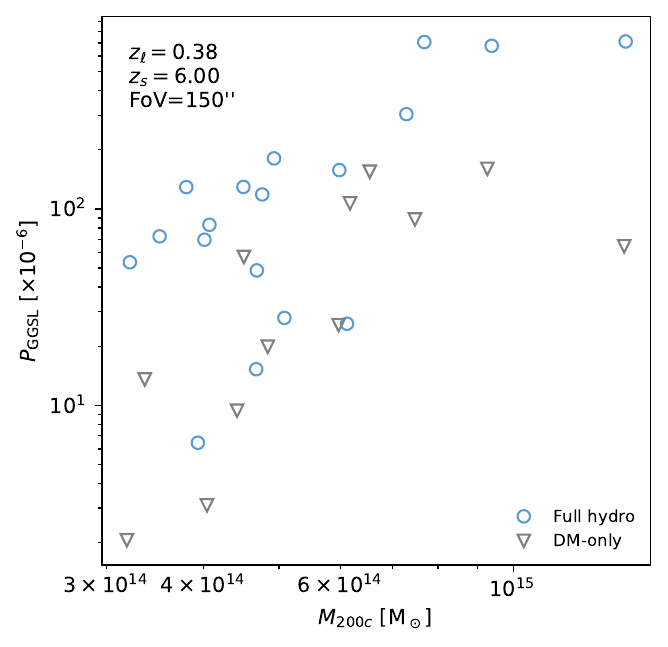}
    \caption{Systematics tests for Galaxy-Galaxy Strong Lensing (GGSL) probability measurements in TNG300-1. (\emph{Left}) The impact on the GGSL probability when including correlated structure (top) and full light cone information (bottom) for a galaxy cluster lens, relative to using only the friends-of-friends identified group as the lens. Subscript ``c'' means Cluster, ``cc'' means Cluster + Correlated, and ``lc'' means Light Cone. (\emph{Right}) Effect of including baryonic physics on the GGSL measurement, using the cluster only as the lens. All groups above $M_{200c}=3\times 10^{14}\,{\rm M_\odot}$ in the group catalog at $z_\ell=0.38$ are used in each panel, and the $\hat{z}$ axis is used as the projection axis in all cases.}
    \label{fig:systematics}
\end{figure*}

\section{Results}\label{sec:results}
\subsection{Primary Lens Properties}\label{sec:results_primary}
Here we examine the dependence of the GGSL probability on the lensing efficiency of the host cluster. The properties of the host can be of significant importance to the measured GGSL cross section, since a larger environmental convergence can boost subcritical galaxy members to criticality or raise their Einstein radii above the lower $\theta_E$ cut in the GGSL cross section definition (Eq.~\ref{eq:GGSL}). In Fig.~\ref{fig:mass_dep} (\emph{left}) we show the GGSL probabilities at $z_s=6$ for both the observed lens models (Sec.~\ref{sec:obs_data}) and simulated data (Sec.~\ref{sec:sim_data}). The simulations utilize only the cluster group for the mass map and do not include correlated structure or uncorrelated structure along the line of sight. The GGSL probability for some clusters is effectively zero, and these objects do not appear on the plot (4 objects of 18 in TNG300-1, for example), though such clusters are likely not comparable to lensing-selected observations.\\

The most apparent feature of this comparison is that the simulations have not produced many clusters with comparable Einstein radii to those of clusters used in the literature for GGSL analysis, meaning the observed and simulated data inhabit largely distinct parts of the $P_{\rm GGSL}-\theta_E$ parameter space. This is somewhat unsurprising, as the observed clusters were selected for being extraordinary lenses, whereas the simulations include all cluster scale objects above a minimum mass cut. The simulated data also utilize the $z$-axis as the projection axis for all objects, regardless of the favorability of that projection. Additionally, since the number of clusters at very large masses is relatively small, if the $z$-axis happens to be an unfavorable projection for the largest objects on average, we will not see the simulations populate the large-$\theta_E$ region of the parameter space. We also show approximate power law fits to each simulation dataset of the form $P_{\rm GGSL}(\theta_E) \propto (\theta_E/\,\qty{10}{\arcsecond})^b$, and find that the scaling of $P_{\rm GGSL}$ with $\theta_E$ is consistent between the TNG300 and TNG-Cluster simulated datasets ($b\simeq 2$), though there are an insufficient number of points to confidently establish a similar power law index for the observations. This simulation power law index may be slightly underestimated due to the fact that objects with zero GGSL probability do not contribute to the fit, and such objects tend to have smaller Einstein radii. \\

To estimate the consistency of the observed and simulated datasets on more equal footing, for each cluster we find the axis among 100 random projections, sampled uniformly on the sphere, that maximizes $\theta_E$ for the TNG-Cluster objects. We show the corresponding GGSL probabilities for these efficient lenses in Fig.~\ref{fig:mass_dep} (\emph{right}). Here it can be seen that a significant number of simulated objects overlap with the parameter space of the observations, enabling a better matched comparison between the datasets which we will address in more detail in Sec.~\ref{sec:results_los}. We also see that some simulated clusters reach Einstein radii of up to $\sim \qty{50}{\arcsecond}$. The number of clusters in the universe with Einstein radii above \qty{40}{\arcsec} is expected to be on the order of $10^1$ to $10^2$ at $z_s=6$ \citep{oguri2009:largest_einstein_radius}. There are no such objects in this simulated dataset when using effectively random lines of sight, but we find that optimizing the projection axis produces $\sim 5$ such large-$\theta_E$ clusters, which may be useful for future comparisons with very efficient lensing clusters such as El Gordo \citep{jee:2014:elgordo} or PLCK-G287 \citep{zitrin2017:plckg287_strong_lens_model}. Note, however, that \cite{oguri2009:largest_einstein_radius} make this prediction with the WMAP1 ($\Omega_m=0.270$, $\sigma_8=0.900$, \citealt{spergel2003:WMAP1}), WMAP3 ($\Omega_m=0.238$, $\sigma_8=0.761$, \citealt{spergel2007:WMAP3}), and WMAP5 ($\Omega_m=0.258$, $\sigma_8=0.796$, \citealt{komatsu2009:WMAP5}) cosmologies, whereas the TNG simulations utilize the Planck 2015 cosmology ($\Omega_m=0.3089$, $\sigma_8=0.8159$, \citealt{planck2016}), and it has been demonstrated that the number of cluster lenses depends strongly on both $\Omega_m$ and $\sigma_8$ \citep{li2006:arc_statistics}.\\

Since all simulated clusters have thus far been placed at $z_\ell=0.38$ to approximately match the median of the observed cluster redshifts, we also examine the evolution of the GGSL probability with lens redshift. Two effects play a role in shaping the lensing properties of the cluster members here; first, the geometric effect of moving the primary lens for a fixed source plane redshift, and second, any evolutionary differences in the cluster properties or distribution and density profiles of cluster members at different epochs. We show the distribution of GGSL probabilities as a function of $z_\ell$, binned by mass, in Fig.~\ref{fig:sigma_vs_zl}. Here we see that despite the fact that critical curves decrease in size as $z_\ell$ increases for fixed $z_s$, the GGSL cross section actually increases as a function of $z_\ell$ in each mass bin. This is consistent with the findings of M23 for $z_s=3$ and $z_s=6$, where the lens-plane redshift evolution was tested for the whole dataset rather than being broken down into mass bins. \\

There are some observed clusters (MACS J1206, Abell 2744 and PSZ1 G311) using the member-boundary field of view definition which exhibit significantly higher values of $P_{\rm GGSL}$ than the simulated counterparts at similar $\theta_E$, as seen in Fig.~\ref{fig:mass_dep} (\emph{right}). To study what remaining systematics could be contributing to this discrepancy, we next examine the effect of line of sight structure, which has not yet been included in the comparison.

% ----------------------------------------------------------------------
\subsection{The Line of Sight}\label{sec:results_los}

In Fig.~\ref{fig:ggsl_los_dep} we show the critical curve and caustic structure for group 0 in the TNG300-1 group catalog with $z_\ell=0.38,\,z_s=4.0$ in three cases: using only particles associated with the group as the lens, including correlated structure by extracting a shell centered on the cluster from the simulation with depth $\sim \pm35\,{\rm Mpc}$, and finally using a full light cone with $\sim 70$ planes. Note that the primary lens plane in the ``Light Cone'' case is precisely the same as the lens plane used in the ``Cluster+Correlated'' case. We also show the magnification map for each considered distribution of matter in the top row, and color-code the critical curves and caustics by those that contribute and do not contribute to the GGSL probability, i.e., those tangential curves that pass the $\theta_E$ cuts (Eq.~\ref{eq:GGSL}). \\

Immediately obvious from the comparison between these panels is that the inclusion of correlated structure makes only a small difference to the critical structure relative to the cluster only case, with some curves merging with the primary critical curve, some secondaries enlarging, and some new secondaries appearing. More striking is the difference made when including the full line of sight in the light cone, in which case the density of secondary critical structures is significantly enhanced. The increase in the number of critical curves as a function of effective Einstein radius is quantified in R26. Many new structures appear in the outer regions, but we caution that if these curves arise from objects significantly separated in redshift from the cluster, lens models in observations would typically not explicitly model them, and thus they would not contribute to the measured GGSL probability.\\

We also see an example of the potential systematic effect of mismatched $\sigma_{\rm FOV}$ definitions in Fig.~\ref{fig:ggsl_los_dep}. Looking at the middle panel, which contains subhalos likely identified as cluster members in observations, we see a void in the lower left corner. The area of this void traced back to the source plane contributes to $\sigma_{\rm FOV}$, thus decreasing $P_{\rm GGSL}$. The same is true for the top right void to a lesser degree. However, in observations, when selecting a field of view curve encompassing the modeled cluster members, such regions may be excluded, increasing $P_{\rm GGSL}$ relative to the simulations. Quantifying the exact extent of this systematic requires implementing the same field of view restriction in simulations as in the observed data, or better yet, modeling mock strong lensing images in simulations in a manner matched to the observations. \\

We quantify the effect of including correlated or full line of sight structure on the GGSL probability in Fig.~\ref{fig:systematics} (left). Here we label the cluster group-only GGSL probability $P_{\rm GGSL,c}$, the probability including correlated structure as $P_{\rm GGSL,cc}$, and the probability with the full light cone as $P_{\rm GGSL,lc}$. We perform this test on TNG data only, since we require the uniform box for the correlated structure and light cone tests (Sec.~\ref{sec:sim_data}), and therefore there are only 18 objects to consider. As in Fig.~\ref{fig:mass_dep} only 14 of these appear in the figure due to some cases having effectively $P_{\rm GGSL}=0$, but such objects are not comparable to the observations. Here we see that the effect of correlated structure is to increase the GGSL probability on average, with a larger fractional boost for systems with large cross sections (and therefore likely large Einstein radii and masses). The boost is on the order of $\sim 10\%$ in most cases, with one case having a large fractional increase of $\sim 50\%$, though this object has a small cross section in the ``cluster only'' case, and the addition of only a small number of supercritical correlated objects significantly increases the cross section in a manner unlikely at larger values of $P_{\rm GGSL}$. A boost of up to $\sim 20\%$ is insufficient to overcome the entirety of the discrepancy between observed and simulated datasets in isolation, but it may be one of several biases which, when considered together, significantly reduce the gap. Note that the depth of the cluster in the ``Cluster+Correlated'' case is $\sim 0.02$, but in the observations the mean depth is $\sim 0.03$, so some structure identified as part of the cluster in observations may not even be included in our correlated structure tests. Such material will be present in the light cone case, however. \\

The light cone adds significantly more power, but in contrast, it has a greater effect on smaller objects. This is expected because the uncorrelated line of sight adds a significant number of secondary critical curves, the number of which scales strongly with the field of view and does not preferentially add structure to smaller or larger objects. Therefore, the fractional effect will be larger for smaller or lower-$P_{\rm GGSL}$ cluster lenses. We caution that the exact boost seen is a strong function of the chosen field of view, and again, many such curves would not be included in observational lens models. This is an important point to remember while comparing observations to the light cone versions. Much of the enhancement to $P_{\rm GGSL, lc}$ arises from critical curves that are not at the cluster redshift and hence lens models do not include them in the computation of the GGSL probability.  It is notable that the cross section for galaxy-galaxy strong lensing due to cluster member galaxies appears to be comparable to the cross section for non-members in the same field of view. The presence of the cluster boosts the cross section for other objects along the line of sight, and this environmental dependence will be examined further in \href{Dutra et al. 2026}{Dutra et al. 2026} (in prep).

% ----------------------------------------------------------------------
\subsection{The Relevance of Baryonic Physics}\label{sec:results_baryons}
We examine the effect of the IllustrisTNG feedback model on the GGSL probability, to complement studies of the effect of feedback models in M22 and M23. We show the GGSL probability for samples with a matching mass cut of $M_{200c} \geq 3\times 10^{14}\,{\rm M_\odot}$ in TNG300-1 and TNG300-1 Dark in Fig.~\ref{fig:systematics} (right). The dark matter-only simulations are systematically lower than the hydrodynamical simulations by $\sim 0.5-1$ dex, as the central densities of the cluster member galaxies are not enhanced by the presence of centrally-dominant baryons. Since these simulations are run with similar initial conditions, the largest objects in the box have approximately the same mass, visible on the high-mass side of Fig.~\ref{fig:systematics} (right). The dark matter density profile itself is also typically contracted in the hydrodynamical simulations \citep{wang2020:TNG_density_profiles, roche:2024}, enhancing the effect beyond the simple addition of baryons onto dark matter-only mass distributions. We confirm a strong impact of the inclusion of baryonic physics on the lensing properties of cluster member galaxies, and that only hydrodynamical simulations should be used for comparing the GGSL probability to observations. Notably, this precludes using the DM-only TNG-Cluster cluster parent simulation for tests of correlated and uncorrelated structure. 

\subsection{Status of the GGSL Tension}\label{sec:ggsl_comparison}
\begin{figure}
    \centering
    \includegraphics[width=1.0\linewidth]{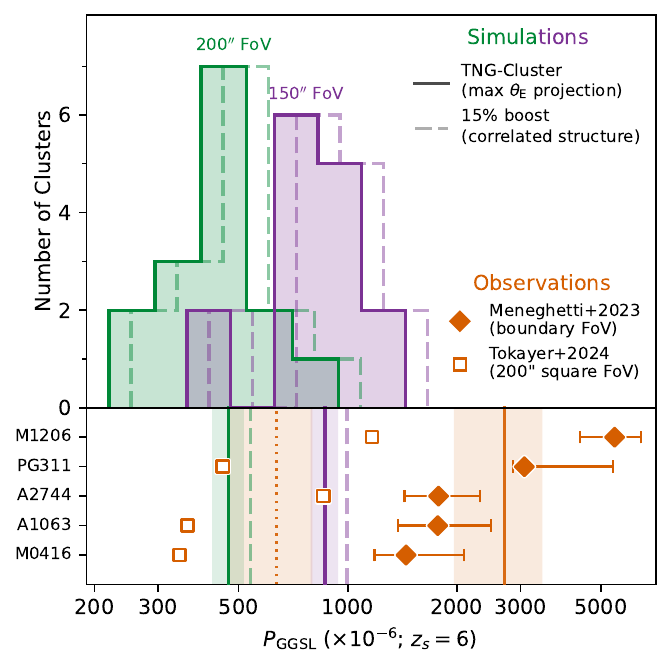}
    \caption{(\emph{Top}) The distribution of Galaxy-Galaxy Strong Lensing Probabilities $P_{\rm GGSL}$ for TNG-Cluster objects, projected along axes maximizing their effective primary Einstein radii with both \qty{150}{\arcsecond} and \qty{200}{\arcsecond} fields of view. A version of each histogram shifted to the right by 15\% is included to approximate conservatively the effect of correlated structure not modelled in the TNG-Cluster data. The simulation data is matched to the window in $\theta_E$ defined by Abell 2744 (\qty{26}{\arcsecond}) and Abell S1063 (\qty{37}{\arcsecond}), and restricted to masses $M_{200c} \geq 10^{15}\,{\rm M_\odot}$. (\emph{Bottom}) The observed clusters with both a \qty{200}{\arcsecond} square boundary (square points) and the boundary curve field of view definition. Also marked are the mean (vertical lines) and standard error on the mean (vertical bands) for each distribution. The standard error is not included for the boosted simulation distributions, and the dotted line is the mean of the square-fov observations.}
    \label{fig:histogram}
\end{figure}

We now attempt to quantify the status of the GGSL tension, accounting for the effect of line of sight structure and using samples matched in mass and Einstein radius. In Fig.~\ref{fig:histogram} we show $P_{\rm GGSL}$ for the max $\theta_E$ projection TNG-cluster objects with $M_{200c}\geq 10^{15}\,{\rm M_\odot}$ to approximately match the observations (Table~\ref{tab:obs}), and in a $\theta_E$ window from \qty{26}{\arcsecond} (Abell 2744) to \qty{37}{\arcsecond} (Abell S1063). We show histograms for both \qty{150}{\arcsecond} and \qty{200}{\arcsecond} fields of view in the simulations, to be comparable to the M23 data and T24 data, respectively. The bins in $P_{\rm GGSL}$ are chosen in each case such that the left edge of the leftmost bin and the right edge of the rightmost bin are set by the min and max $P_{\rm GGSL}$ values in the simulated samples, thus making the footprint of the histogram as small as possible. 
We also show $P_{\rm GGSL}$ for the observations as measured with the boundary curve field of view curve definition of M22 and M23, and with the square field of view definition used in M20, T24 and \cite{dutra2025:ggsl}. 
Again the boundary field of view definition is likely more accurate for the observations than the square field of view, with the most appropriate comparison to simulations lying somewhere between.\\

\begin{figure*}
    \centering
    \includegraphics[width=0.95\linewidth]{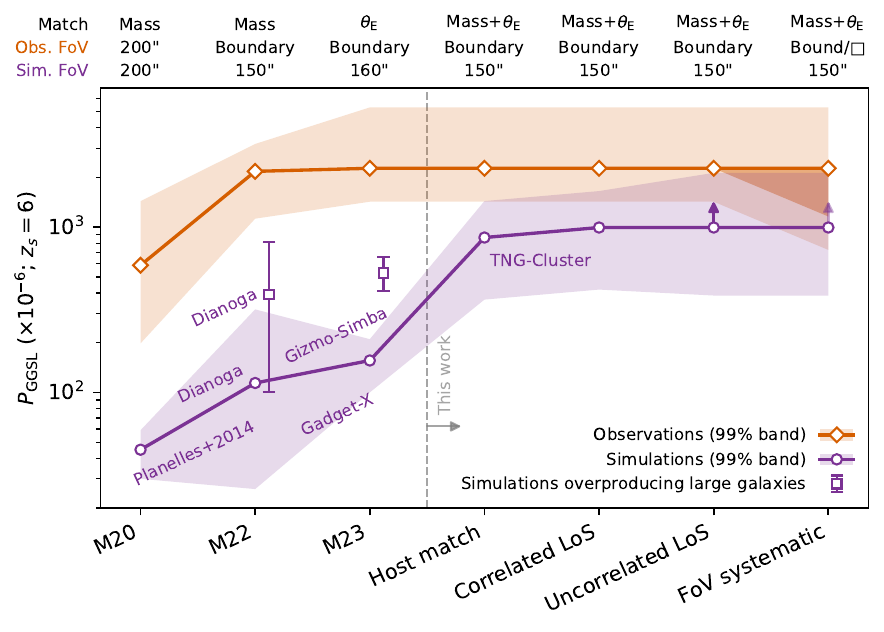}
    \caption{A schematic of the status of the Galaxy-Galaxy Strong Lensing (GGSL) tension over time, as various new datasets were considered and methodological improvements were made. The dark central line in the observations band is the mean due to small sample size (matching M23), while for the simulations it is the median. All points to the right of the gray dotted line come from the data discussed in this paper. The effects to the right of the dashed line are shown cumulatively, so each $P_{\rm GGSL}$ value includes the contribution of those effects to the left. The vertical arrow for ``Uncorrelated LoS'' indicates that the effect likely increases the median, but this has not been quantified. The lower side of the dark region in the ``FoV Systematic'' column marks the mean $P_{\rm GGSL}$ when using square fields of view in the observations rather than the boundary curve definition, approximated by scaling the T24 observations mean by the ratio of the median \qty{200}{\arcsecond} and \qty{150}{\arcsecond} $P_{\rm GGSL}$ values in the mass- and $\theta_E$-matched simulation distributions. The low side of the observations band in that column has been reduced to a fraction below the T24 mean equal to the fraction it would otherwise fall below the M23 mean. Since the GGSL probability scales inversely with the source plane field of view area, the field of view for each measurement is also noted above the plot area. Care should be taken comparing points horizontally with differing fields of view, but vertical comparisons reflect the status of the tension at that stage in time/development.} 
    \label{fig:tension}
\end{figure*}

In Fig.~\ref{fig:histogram} we observe that despite being matched in mass and Einstein radius, a clear offset exists between the simulation distributions and their field of view-matched counterparts in the observations. In the lower panel, the means and standard error on the means for each distribution are shown, where it is apparent that the boundary-curve observations lie significantly above the simulations, and the square field of view observations are approximately consistent with the \qty{200}{\arcsecond} simulations. The boosted simulation distributions approximate the effect of correlated structure in this sample using the small number of objects at high $P_{\rm GGSL}$ in TNG300-1, which experience a boost in GGSL probability of $\sim 15\%$. Even comparing the means of the boosted distributions with the M23 data the offset remains clear. Note that the horizontal axis is logarithmic and that the magnitude of this discrepancy is large. Note also that since the observations exhibit a spread in $z_\ell$ and the simulations all use $z_\ell=0.38$, an additional scatter of a similar magnitude to the correlated structure boost is present in the observations but not simulations (see Fig.~\ref{fig:sigma_vs_zl}).\\

\begin{figure}
    \centering
    \includegraphics[width=1.0\linewidth]{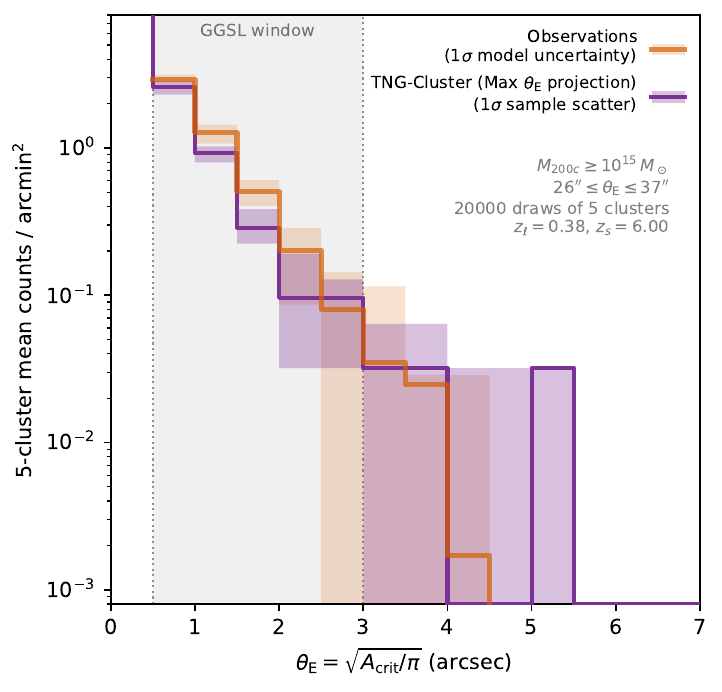}
    \caption{The distribution of effective Einstein radii for critical curves in the matched observed and simulated datasets, averaged over the 5 considered observations, and samples of 5 clusters drawn without replacement from the matched simulations. The simulation data here includes no contribution from correlated or uncorrelated structure, and uses a \qty{150}{\arcsecond} field of view. The simulations band is based on the variability in each bin due to different samples of clusters, whereas the observations band is based upon uncertainty in the strong lens model of these objects.}
    \label{fig:critical_radius_dist}
\end{figure}

We also demonstrate the evolution of the tension with different datasets and the various stages in methodological development in Fig.~\ref{fig:tension}. Here we represent the difference between the observed and simulated datasets as a function of previous work (\emph{left}) and the systematics discussed in this paper (\emph{right}). A number of caveats for this figure are important to note. 
\begin{enumerate}[
  align=left,
  leftmargin=\parindent,
  itemindent=0pt,
  labelsep=0.5pt,
  labelwidth=1em,
  itemsep=0.3em
]
    \item The datasets used, observed, simulated, or both, differ for each of the M20, M22, M23, and ``This work'' portions of the plot. The simulation names are annotated, and in the case of M20 the \citep{planelles2014:m20_simulations} simulations were used for the theoretical predictions. For the observations the same objects are frequently shared between columns, but in some cases with different field of view definition or specific lens models.
    \item The field of view definition changes across datasets, in particular for the simulations, which has a direct impact on the scale of $P_{\rm GGSL}$. 
    \item The ``Host match'' column in this work is not the first time that the Einstein radius has been used to match the observed and simulated samples, as noted by the Match column for M23, but is the first time that both the Einstein radius and mass were simultaneously explicitly matched. The cluster mass distribution in M23 is likely not poorly matched to the observations, as the 324 most massive objects from a $1\,{\rm Gpc}$ box are used in that analysis, but TNG-Cluster also utilizes a $1\,{\rm Gpc}$ box from which we find only 15 matches (these matched clusters are the simulation data presented in Fig.~\ref{fig:histogram} and Fig.~\ref{fig:tension}).
    \item The $\theta_E,{\rm max}$ used as the upper limit in the GGSL definition (Eq.~\ref{eq:GGSL}) is not consistent across the plot, with $\theta_E,{\rm max}=\qty{5}{\arcsecond}$ in M20, and \qty{3}{\arcsecond} elsewhere. 
    \item The magnitude of the effect of uncorrelated structure is not known, and is difficult to quantify without performing lens modeling on simulated data in a manner matched to the observations. This effect is approximated here as an increase in the scatter by $20\%$, motivated by the scatter introduced in total cluster critical area within \qty{100}{\arcsecond} of the cluster potential minimum found in R26, but this approximation is crude. Since this effect is also likely to introduce additional galaxy-scale objects that are not in our correlated volume but may be considered as cluster galaxies by the redshift ranges used in observations (Table~\ref{tab:obs}), we also display an upward arrow indicating that this effect is likely to increase the simulation median (or equivalently, decrease the probability in the observations if accounted for).
\end{enumerate}

This means data points cannot be compared on exactly equal footing across columns. Instead, this visualization primarily attempts to demonstrate schematically the status of the tension at each stage and how that tension evolves with different methodologies and data. \\

What seems to be suggested by this figure is that matching the host properties is very important for the consistency of the two datasets. Matching only in $\theta_E$ or in mass seems to be insufficient, and the nature of the feedback model implemented in the simulations is undoubtedly highly impactful. This is made clear by the fact that the max-projection TNG-Cluster data matched in $\theta_E$ but not mass to the observations in Fig.~\ref{fig:mass_dep} is in lower tension with the M23 observations than the $\theta_E$-matched simulations of M23. We show as individual errorbars the simulation distributions from M22 and M23 which are computed using simulation data that was demonstrated to overproduce massive galaxies relative to the observations. In these cases the feedback model produces higher-$P_{\rm GGSL}$ clusters more in line with observations, at the expense of producing galaxies which do not appear to have observational counterparts. It is therefore reasonable to ask if the relative success of the TNG-Cluster data is paired with a similar discrepancy in the distribution of Einstein radii.\\

In Fig.~\ref{fig:critical_radius_dist} we produce the same test as performed in M22 and M23, which is to examine the average number of critical curves per square arcminute binned in effective Einstein radius. In observations this represents an average across the whole sample, but in simulations the appropriate comparison is to the collection of curves obtained by, for each curve, sampling 5 objects randomly from the mass- and $\theta_E$-matched clusters, computing the mean counts per arcminute in each bin across that sample of 5, and obtaining many such 5-object means per bin. We produce 20000 such curves in the simulation data, and show the median and $16^{\rm th}$/$84^{\rm th}$ percentiles of these curves as a line with a band. The simulation bands thus represent a variance due to sampling different objects, whereas the observed bands arise from utilizing different samples in the Markov chain used to constrain the lens model, and can thus be considered statistical uncertainty in the lens models. Here we see a deficit of $\theta_E\simeq \qty{1.5}{\arcsecond}$ objects by a factor of up to $\sim 2$, consistent with Gizmo-Simba clusters from M23, and 10xB20 clusters from M22. Note however that the boundary curve field of view definition could bias the observations curve high, as regions without galaxies in the outskirts will not reduce the mean counts per area in that case. Both the Gizmo-Simba and 10xB20 simulations are those marked as overproducing large galaxies in Fig.~\ref{fig:tension}. However, beyond \qty{4}{\arcsecond} we find that for 5-object samples, the simulation histogram is everywhere consistent with zero, indicating that massive galaxies are not significantly overproduced in TNG-Cluster. This said, larger-$\theta_E$ objects do exist in the simulated clusters. A simple figure for comparison is that in 27\% of the 20000 draws of 5 clusters without replacement from the matched clusters, there were no clusters with a secondary critical curve above \qty{4.5}{\arcsecond} (the largest $\theta_E$ in the observed cluster members). This does not represent a confident tension between the datasets. Note also that the small sample of matched clusters from which we draw samples of 5 (15 total clusters) limits the robustness of this test, as many samples of 5 will be very close or the same as others. A further complication is that only the simulation group is used as the lens in this comparison, including no correlated structure or the effect of uncorrelated line of sight material. This may resolve the deficit in simulations at low-$\theta_E$, or create a discrepancy at large-$\theta_E$ as seen in M22 and M23, though these studies did not include correlated structure in the computations.

\section{Conclusions}\label{sec:conclusions}
In this paper we have studied the established tension in the lensing efficiency of galaxy cluster member galaxies between a) lens models of observed strong lensing clusters, and b) ray-traced mass maps created from clusters and their environments in cosmological simulations. This tension is quantified by the Galaxy-Galaxy Strong Lensing (GGSL) probability $P_{\rm GGSL}$ (Eq.~\ref{eq:pggsl}). We have focused in particular on the dependence of the GGSL probability on the properties of the host cluster, namely the lensing efficiency of the large-scale halo quantified by the primary critical curve effective Einstein radius $\theta_E$, and the redshift of the cluster $z_\ell$. We also study the inclusion of baryonic physics and systematic effects of the line of sight in simulated data, comparing the inferred GGSL probability when only the simulated cluster is used, or when including correlated and/or uncorrelated structure along the line of sight. We make the following observations:
\begin{itemize}[
  align=left,
  leftmargin=\parindent,
  itemindent=0pt,
  labelsep=0.5pt,
  labelwidth=1em,
  itemsep=0.3em
]
\item The GGSL probability scales strongly with the Einstein radius of the host cluster, approximately $P_{\rm GGSL} \propto \theta_E\,^{2}$ but with significant scatter on the order of $\pm 0.5$ dex in highly-sampled regimes.  Matching the simulated data to the observations in mass and Einstein radius, the latter by choosing favorable lines of sight for the simulated clusters, brings the observed and simulated datasets into closer agreement than previously reported. However, a persistent deficit in the simulation GGSL probability of roughly a factor of $\sim 2 - 3$ between $\theta_E=\qty{26}{\arcsecond}$ and $\theta_E=\qty{37}{\arcsecond}$ remains when using the cluster member-bounding curve field of view definition of M22 and M23.
\item The field of view definition is highly impactful for the inferred GGSL probabilities, and the simulated data is not in significant tension with the observations with square fields of view. The square field of view certainly underestimates the probability, while the cluster member boundary definition can overestimate $P_{\rm GGSL}$ depending on the large-scale properties of the cluster such as ellipticity in the plane of the sky, in addition to the methods used to select cluster members.
\item Correlated line of sight structure, modeled here as material within $\sim 35\,{\rm cMpc}$ of the cluster plane along the line of sight, enhances the GGSL probability on average, with a larger fractional effect for objects with large GGSL probabilities. This is accomplished by both enhancing the sizes of small critical curves into the window used to compute $P_{\rm GGSL}$ (and rarely, out of that window), and by introducing additional supercritical objects which would likely be identified as cluster members in observations. This accounts for a $\sim 15\%$ boost to $P_{\rm GGSL}$ over using the friends-of-friends identified group only for the largest objects in the simulation, but such objects are few in number.
\item Uncorrelated structure along the line of sight introduces a significant boost to $P_{\rm GGSL}$ for all systems, with the largest effect (factor of $\sim 2$) for clusters with low-$P_{\rm GGSL}$ in isolation, and a diminished effect (an increase of $\sim 20\%$) for the systems with the highest $P_{\rm GGSL}$ in isolation. This is not directly comparable to inferred GGSL probability for observations, however, because only those galaxies identified as cluster members contribute to $P_{\rm GGSL}$ in the lens model. Interestingly, we find that a galaxy-galaxy lens in a cluster field (in a $\qty{150}{\arcsecond}\times \qty{150}{\arcsecond}$ square field of view) is approximately equally likely to be a cluster member as it is to be located elsewhere along the line of sight.  \\
\end{itemize}

However, this comparison remains subject to significant limitations, and to robustly establish the magnitude of this tension a number of major systematics must be addressed. First, the selection function of the observations used must be well-understood, and reproduced in the simulations. This is difficult because only clusters with robust lens models including stellar kinematics for the cluster members can be used, since the cluster member properties and large-scale mass distribution must be simultaneously well-constrained to make this measurement. 
Second, the methods of measuring $P_{\rm GGSL}$, including the field of view definition, must be matched in the observations and simulations. The mass distribution in observations is constrained through the standard techniques of cluster strong lens modeling, but in simulations the mass distribution is measured directly from the simulation particles and cells. Ideally, mock strong lensing images with a matched selection function to the observations should be modeled in a manner comparable to the observations, such that the systematics of strong lensing mass reconstruction are reproduced in the simulation data. This is difficult both because generating such mocks from the simulation data is very challenging, and because once those mocks exist they must be modeled, a process which is highly specialized and demanding on cluster scales. In R26 we took the first step and introduced a pipeline to generate such simulation mocks. Overcoming the remaining barriers of matching the selection function and performing methodologically-matched lens modeling in the simulations would place the comparison between observed and simulated GGSL probabilities on firmer ground.

% ------------ ACKNOWLEDGEMENTS ------------
\section*{Acknowledgments}
% ------------ SOFTWARE ------------
This work makes use of the following software: \textsc{Python} \citep{python}, 
\textsc{numpy} \citep{numpy:2020}, \textsc{scipy} \citep{scipy:2020}, 
\textsc{astropy} \citep{astropy:2013, astropy:2018}, 
\textsc{jupyter} \citep{jupyter}.

% ------------ BIBLIOGRAPHY ------------
% \bibliography{bibliography}{}
% \bibliographystyle{aasjournal}
\bibliographystyle{aasjournal}
\bibliography{bibliography}

% ---------------- APPENDIX ----------------
% \begin{appendix}
% \input{content/appendix.tex}
% \end{appendix}

\end{document}